\documentclass[11pt,fleqn]{article}       
\usepackage{geometry}
\usepackage[T1]{fontenc}
\usepackage[utf8]{inputenc}
\usepackage{authblk}
\usepackage{mathrsfs}
\usepackage[font=small,labelfont=bf]{caption}
\usepackage{graphicx}
\usepackage{multirow}
\usepackage{amsmath,amssymb,amsfonts}
\usepackage{verbatim}
\usepackage{bm}
\usepackage{color}
\usepackage{ulem}
\usepackage{mathtools}
\usepackage{cite}
\usepackage{colortbl}
\usepackage{cancel}
\definecolor{light-gray}{gray}{0.85}
\usepackage[pdftex,colorlinks,pdfpagelabels]{hyperref}
\hypersetup{
   bookmarksnumbered,
   citecolor={blue},
   linkcolor={blue},
   urlcolor={blue},
   filecolor={blue}
} 

\newcommand{\AddrTexas}{%
\textit{Department of Physics, The University of Texas at Austin, Austin, 78712 TX, USA}
}
\newcommand{\AddrStockholm}{
\textit{Oskar Klein Center for Cosmoparticle Physics, University of Stockholm, 10691 Stockholm, Sweden}
}
\newcommand{\AddrNordita}{
\textit{Nordita, KTH Royal Institute of Technology and Stockholm University, 10691 Stockholm, Sweden}
}

\usepackage{fancyhdr}
 \usepackage[bottom]{footmisc}
\fancypagestyle{plain}{%
    \fancyhead[R]{NORDITA-2024-002, UT-WI-02-2024}
    
}
\date{}
\title{Origin of the Stochastic Gravitational Wave Background: First-Order Phase Transition vs. Black Hole Mergers
}
\author[1]{Martin Wolfgang Winkler\thanks{martin.wolfgang.winkler@gmail.com}}
\author[1,2,3]{Katherine Freese\thanks{ktfreese@utexas.edu}}
\affil[1]{\AddrTexas}
\affil[2]{\AddrStockholm}
\affil[3]{\AddrNordita}

\begin{document}
\maketitle
\vspace*{0mm}
\begin{abstract}
The NANOGrav, Parkes and European Pulsar Timing Array (PTA) experiments have collected strong evidence for a stochastic gravitational wave background in the nHz-frequency band. In this work we perform a detailed statistical analysis of the signal in order to elucidate its physical origin. Specifically, we test the standard explanation in terms of supermassive black hole mergers against the prominent alternative explanation in terms of a first-order phase transition. By means of a frequentist hypothesis test we find that the observed gravitational wave spectrum prefers a first-order phase transition at $2-3\sigma$ significance compared to black hole mergers (depending on the underlying black hole model). 
This mild preference is linked to the relatively large amplitude of the observed gravitational wave signal (above the typical expectation of black hole models) and to its spectral shape (which slightly favors the phase-transition spectrum over the predominantly single power-law spectrum predicted in black hole models).
The best fit to the combined PTA data set is obtained for a phase transition which dominantly produces the gravitational wave signal by bubble collisions (rather than by sound waves).  The best-fit (energy-density) spectrum features, within the frequency band of the PTA experiments, a crossover from a steeply rising power law (causality tail) to a softly rising power law; the peak frequency then falls slightly above the PTA-measured range. Such a spectrum can be obtained for a strong first-order phase transition in the thick-wall regime of vacuum tunneling which reheats the Universe to a temperature of $T_*\sim \text{GeV}$. A dark sector phase transition at the GeV-scale provides a comparably good fit. 
\end{abstract}
\clearpage

\section{Introduction}
The North American Nanohertz Observatory for Gravitational Waves (NANOGrav)~\cite{NANOGrav:2020bcs,NANOGrav:2023gor}, the European Pulsar Timing Array (EPTA)~\cite{Chen:2021rqp,EPTA:2023fyk}, the Parkes Pulsar Timing Array (PPTA)~\cite{Goncharov:2021oub,Reardon:2023gzh} and the Chinese Pulsar Timing Array~\cite{Xu:2023wog} have recently announced the detection of a low-frequency stochastic process which affects pulsar timing residuals. Within their latest data release NANOGrav, EPTA and PPTA, furthermore, reported strong evidence for quadrupolar Hellings-Downs correlations~\cite{Hellings:1983fr} which are characteristic for gravitational wave (GW) sources~\cite{NANOGrav:2023gor,NANOGrav:2023hvm,EPTA:2023fyk,EPTA:2023xxk,Reardon:2023gzh}. The Pulsar Timing Arrays (PTAs) have, thus, very likely discovered a stochastic GW background in the nHz-frequency band. 

The canonical explanation for such a low-frequency background consists in the joined GW emission from a large population of merging supermassive black holes (SMBHs)~\cite{Haehnelt:1994wt,Rajagopal:1994zj,Jaffe:2002rt,Wyithe:2002ep,Sesana:2004sp,Sesana:2008mz,Burke-Spolaor:2018bvk,Middleton:2020asl}. SMBHs, which commonly exist in the centers of galaxies, can form binary systems when two galaxies merge. When they are sufficiently close the binary black holes quickly transfer orbital energy into GWs. This makes their distance further shrink such that they finally coalesce. However, in order for GW emission to become efficient, black hole separations as low as $0.1 - 0.001\:\text{pc}$ must first be reached~\cite{Begelman:1980vb}. The absence of established physical processes which would remove orbital energy from SMBH binaries at relative distances of 0.1 to 1$\:$pc -- when encounters with stars (in the simplest scenarios) tend to become inefficient -- is known as the ``final-parsec problem''~\cite{Milosavljevic:2002ht}. While attempts to mitigate the final-parsec problem exist -- invoking for instance more efficient stellar relaxation mechanisms~\cite{Quinlan:1997qe,Yu:2001xp,Zhao:2001py,Milosavljevic:2002bn,Merritt:2003pf,Berczik:2006tz}; the formation of an accretion disc which could drain orbital energy from the SMBHs at sub-kpc separations~\cite{Begelman:1980vb,Gould:1999ia}; physics from beyond the Standard Model~\cite{Koo:2023gfm,Bromley:2023yfi}; or other mechanisms briefly discussed in Sec.~\ref{sec:gwbhm} below -- these ideas require further scrutiny by means of simulations. At present, the magnitude (if any) of the GW signal from SMBH mergers, therefore, remains uncertain. 

In this light, several alternative explanations for the observed stochastic GW background have been explored. These include a first-order phase transition ~\cite{Caprini:2010xv,Schwaller:2015tja,Kobakhidze:2017mru,Nakai:2020oit,Addazi:2020zcj,Ratzinger:2020koh,Brandenburg:2021tmp,NANOGrav:2021flc,Borah:2021ocu,DiBari:2021dri,Lewicki:2021xku,Ashoorioon:2022raz,Freese:2022qrl,Freese:2023fcr,Cruz:2023lnq}, a cosmic-string network~\cite{Vilenkin:1981bx,Vachaspati:1984gt,Damour:2004kw,Siemens:2006yp,Olmez:2010bi,Ringeval:2017eww,Ellis:2020ena,Blasi:2020mfx,Buchmuller:2020lbh}, domain walls~\cite{Ferreira:2022zzo,Barman:2023fad,Babichev:2023pbf,Gelmini:2023kvo,Zhang:2023nrs,Kitajima:2023cek,Guo:2023hyp}, inflationary quantum fluctuations (in non-standard inflation models with either an extremely blue tensor spectrum~\cite{Vagnozzi:2020gtf,Kuroyanagi:2020sfw} or a strongly enhanced small-scale scalar power spectrum~\cite{Vaskonen:2020lbd,DeLuca:2020agl,Kohri:2020qqd,Domenech:2020ers}) and ``audible axions''~\cite{Machado:2018nqk,Machado:2019xuc}.

In this work we will test the standard explanation for the stochastic GW background in terms of SMBHs against the alternative explanation in terms of a first-order phase transition (FOPT). We select the latter among the list of alternatives due to its strong theoretical motivation -- a FOPT occurs in many prominent early universe models in the context of inflation~\cite{Adams:1990ds,Linde:1990gz,Lopez:2013mqa,Ashoorioon:2022raz,Freese:2022qrl,Freese:2023szd}, thermal inflation~\cite{Lyth:1995ka,Barreiro:1996dx}, the breaking of a gauge symmetry (see e.g.~\cite{Jaeckel:2016jlh,Jinno:2016knw,Addazi:2017gpt,Hashino:2018zsi,Croon:2018erz,Marzo:2018nov,Breitbach:2018ddu,Baratella:2018pxi,Azatov:2019png,Lewicki:2020azd}), dark matter production~\cite{Chung:1998ua,Kolb:1998ki,Petraki:2011mv,Falkowski:2012fb,Azatov:2021ifm,An:2022toi,Freese:2023fcr} or baryogenesis~\cite{Kuzmin:1985mm,Rubakov:1996vz,Shelton:2010ta,Katz:2016adq,Hall:2019rld} -- and because it can easily accommodate the large amplitude of the observed GW signal (as opposed to some other cosmological explanations)~\cite{NANOGrav:2023hvm}.

After extracting the PTA data we perform an extensive frequentist hypothesis test which investigates the observed spectral properties of the signal against the expectation from both hypotheses. Compared to previous studies on these two hypotheses which relied on a single PTA~\cite{NANOGrav:2023hvm,EPTA:2023xxk,Ellis:2023oxs} (or two~\cite{Figueroa:2023zhu}), our analysis profits from a joined fit to the combined data of NANOGrav, PPTA and EPTA. Furthermore, due to our frequentist approach, we can avoid a strong dependence of our results on prior assumptions on the phase transition parameters -- an issue which has plagued previous Bayesian analyses (see e.g. the discussion in~\cite{Bringmann:2023opz}). We find a preference for the FOPT against the SMBH hypothesis at $2-3\sigma$ significance. Within the subcases of FOPTs we identify a slight preference for a phase transition in which the GW signal is dominated by bubble collisions (as opposed to a signal dominated by acoustic GWs produced during the bubble expansion stage).

This work is organized as follows: in Sec.~\ref{sec:gwbhm} and Sec.~\ref{sec:fopt} we review the SMBH and FOPT hypotheses and introduce the calculation of the associated GW spectra. In Sec.~\ref{sec:signal} we extract the spectral information of the GW signal from the latest NANOGrav, PPTA and EPTA data sets. In Sec.~\ref{sec:test} we perform our main analysis where we test the FOPT hypothesis against the SMBH hypothesis. Finally, Sec.~\ref{sec:conclusion} contains our conclusions.

\section{Gravitational Waves from Black Hole Mergers}\label{sec:gwbhm}

Mergers of SMBHs have long been considered as a possible source of a stochastic GW background in the PTA frequency band~\cite{Haehnelt:1994wt,Rajagopal:1994zj,Jaffe:2002rt,Wyithe:2002ep,Sesana:2004sp,Sesana:2008mz,Burke-Spolaor:2018bvk,Middleton:2020asl}. A large population of SMBH binaries could have formed from the central black holes of merging galaxies. When the binary black holes come very close, they coalesce because GWs significantly drain their orbital energy. The joined emission of many merging SMBHs is considered as the standard explanation of the GW signal observed by the PTA experiments.

However, the occurrence of SMBH mergers is far from trivial. While binary SMBHs can efficiently shrink their separations by close encounters with stars, this process should stall at parsec-distances when almost all stars along their orbit have been ejected (unless there exists an efficient mechanism which refills the loss cone with stars). GW emission, on the other hand, only becomes efficient at separations of  $0.1 - 0.001\:\text{pc}$~\cite{Begelman:1980vb}. The absence of established physical processes which would remove orbital energy from SMBH binaries at relative distances of 0.1 to 1$\:$pc is known as the ``final-parsec problem''~\cite{Milosavljevic:2002ht}. Only if the final-parsec problem is solved, SMBH mergers present a possible source of measurable gravitational radiation.

Attempts to mitigate the problem include the introduction of more efficient stellar-relaxation mechanisms~\cite{Quinlan:1997qe,Yu:2001xp,Zhao:2001py,Milosavljevic:2002bn,Merritt:2003pf,Berczik:2006tz} as well as the consideration of triple-SMBH interactions~\cite{Blaes:2002cs,Hoffman:2006iq,Amaro-Seoane:2009ucl} (although the requirement of a third black hole would likely reduce the number of mergers). Another possibility is that merging galaxies (which host the SMBHs) drive gas inflow towards the galactic centers leading to the formation of an accretion disk around the black hole binaries~\cite{Begelman:1980vb,Gould:1999ia}. The interaction with the accretion disk could then inflict additional friction on the orbiting black holes when stellar dynamics become insufficient such that the black holes ultimately merge -- a theory which is, however, not fully supported by the most recent hydrodynamical simulations (see~\cite{Lai:2022ylu}). Another possibility to inflict additional friction on the SMBHs consists in invoking a halo of ultralight dark matter~\cite{Koo:2023gfm,Bromley:2023yfi}.

Assuming that SMBHs ultimately merge the total GW signal is obtained by integrating the GW emission of individual binaries throughout the Universe (see e.g.\cite{Phinney:2001di}). The calculation requires a modeling of galaxy masses and merger rates, a mass relationship between host galaxies and SMBHs as well as a binary evolution prescription. The latter poses a particular challenge as long as the physics of the final parsec is not resolved. A common approach in the literature is to assume that some (known or unknown) interaction with the astrophysical environment effectively shrinks the binary separation (=``hardens the binary'') until GW emission becomes efficient. The hardening timescale is parameterized based on phenomenological hardening models, hydrodynamical simulations or empirical arguments -- an approach which, however, has its limitations in the absence of a satisfactory solution to the final-parsec problem.

In the simplest case in which the SMBH binaries (i) are distributed continuously in mass, distance and orbiting frequency, (ii) move on circular orbits and (iii) evolve purely due to GW emission\footnote{This condition applies to the SMBH binaries which dominate the GW signal, i.e.\ those at very small separation. In order to reach the small separations a mechanism to overcome the final-parsec problem is still required.}, the GW spectrum expressed in terms of the critical density is given by,
\begin{equation}\label{eq:omegahc}
 \Omega_{\text{GW}}^{\text{SMBH}} (f) = \frac{2\pi^2}{3 H_0^2}\,A_{\text{GW}}^2\, f^2   \:\left(\frac{f}{f_0}\right)^{3-\gamma_{\text{GW}}}\,,
\end{equation}
where, by convention, the reference frequency is picked as $f_0=1\:\text{yr}^{-1}$. The GW amplitude and power-law index read~\cite{Phinney:2001di},
\begin{align}\label{eq:referencemodel}
A_{\text{GW}}^2 &= \frac{4\pi}{3c^2}(2\pi f_0)^{-4/3}\int dM \,d(\tfrac{m_2}{m_1}) \,dz\, \frac{\partial^3\eta}{\partial M\, \partial(\tfrac{m_2}{m_1})  \,\partial z} \frac{(G\mathcal{M})^{5/3}}{(1+z)^{1/3}}\,,\nonumber\\
\gamma_{\text{GW}} &= \frac{13}{3}\,,
\end{align}
where $m_{1,2}$ denote the black hole masses, $M=m_1+m_2$ the total mass of the binary, $\mathcal{M}=(m_1 m_2)^{3/5}/M^{1/5}$ the chirp mass and $z$ the emission redshift. Furthermore, $\eta$ stands for the comoving volumetric number density of binaries.

The above power-law parameterization with a fixed spectral index of $\gamma_{\text{GW}}=13/3$ has long been considered the standard reference model for SMBH-induced GW emission. The amplitude is either taken as a free parameter or derived from Eq.~\eqref{eq:referencemodel} (based on a modeling of the SMBH population). Early analyses by the NANOGrav, PPTA and EPTA collaborations tested their data against this reference model~\cite{NANOGrav:2020bcs,Chen:2021rqp,Goncharov:2021oub}. However, within recent years, progress towards a more realistic description of SMBH mergers has been made. 

The following three effects cause deviations of the GW spectrum from the reference prediction (see~\cite{NANOGrav:2023hfp} for a summary): (i) while a large number of SMBHs contributes to the GW signal at low frequency $f\ll 1\:\text{yr}^{-1}$, the high-frequency part is generated by a handful of binaries. The discreteness of the binary population results in a reduction of GW power at high frequency $f\gtrsim 1\:\text{yr}^{-1}$ compared to the prediction from a continuous distribution (see e.g.~\cite{Sesana:2008mz}). (ii) Non-vanishing eccentricities of the binary orbits move GW energy from lower to higher frequencies due to the emission of GWs at higher harmonics (while circular binaries strictly emit GWs at twice the orbital frequency)~\cite{Peters:1963ux,Enoki:2006kj,Sesana:2010qb}. (iii)
Binary interactions with the astrophysical environment -- which are required to solve the final-parsec problem -- tend to speed up the binary evolution. Compared to a purely gravitational-wave driven evolution this leads to a depletion of the GW spectrum  at low frequency because a fraction of the orbital energy is injected into the environment (rather than being converted to gravitational radiation)~\cite{Begelman:1980vb,Kocsis:2010xa}. 
Whether this crossover occurs below or within the frequency band of PTAs depends on the (unknown) physics of the final parsec.

In this work we will employ two black hole models which were recently constructed for the interpretation of PTA signals:
\begin{itemize}
\item model 1 (introduced by the NANOGrav collaboration in~\cite{NANOGrav:2023hvm}) is based on a simulation of realistic SMBH populations performed with the holodeck package. The binary-evolution is purely gravitational-wave-driven and orbital eccentricities are neglected. Deviations from the reference model (with fixed $\gamma_{\text{GW}}=13/3$) still arise due to the discretization of the SMBH population which accounts for the Poissonian fluctuations in the true number of binaries in any given spatial volume.
\item model 2 (developed by the EPTA collaboration in~\cite{EPTA:2023xxk}) employs a subset of SMBH populations which were constructed from estimated galaxy merger rates and empirical black hole-host relations in~\cite{Sesana:2012ak,Rosado:2015epa}. The subset accounts for an upward revision of the SMBH-galaxy relations~\cite{McConnell:2011mu} and -- motivated by hydrodynamical simulations~\cite{Farris:2013uqa,Capelo:2016yrq} -- focuses on binaries with accretion (dominantly onto the less massive black hole) prior to merger. The initial eccentricities $e_0$ of the binaries are varied between 0 and 1 and their subsequent evolution is tracked using the hardening models of~\cite{Sesana:2010qb}. These account for the combined effects of stellar hardening and GW emission for binaries coupled to their astrophysical environment.
\end{itemize}
Another black hole model which incorporates environmental effects has been provided by the NANOGrav collaboration in~\cite{NANOGrav:2023hfp}. The latter is not included in our analysis because it yields a similar fit as model~2 introduced above.

We note that the preselection of binary populations in model 2 increases the predicted mean gravitational amplitude compared to model 1. Furthermore, by including eccentricities and environmental effects, model 2 produces larger deviations from the reference model (Eq.~\eqref{eq:referencemodel}) compared to model 1. However, given the lack of eccentric binary black hole detections at interferometers~\cite{LIGOScientific:2023lpe}, it yet needs to be established whether eccentricity plays a role in merging SMBH systems. In this light, the different assumptions in the two black hole models amount to systematic uncertainties which cannot be resolved with present data. By including both models in our analysis we aim at reflecting the uncertainty associated with the underlying SMBH binary description.

When implementing the black hole models we follow~\cite{NANOGrav:2023hvm} and approximate the resulting GW spectra by the power-law form of Eq.~\eqref{eq:omegahc} (without imposing $\gamma_{\text{GW}} = 13/3$ as in the reference model~\cite{Phinney:2001di}). This effectively amounts to averaging over small tilts in the GW spectra induced by the discreteness of binary populations, eccentricities of the orbits and environmental effects -- a procedure which is justified by the relatively narrow bandwidth of PTAs~\cite{NANOGrav:2023hvm}. The probability distributions in the ($A_{\text{GW}}$,$\gamma_{\text{GW}}$)-plane are extracted from~\cite{NANOGrav:2023hvm} (for model 1) and from~\cite{EPTA:2023xxk} (for model 2). For simplification, we approximate them by bivariate normal distributions in ($\log_{10}A_{\text{GW}}$,$\gamma_{\text{GW}}$) whose mean $\mu_{\text{GW}}$ and covariance matrix $\sigma_{\text{GW}}$ are given by 
\begin{align}\label{eq:SMBHparameters}
\mu_{\text{GW}} &= (-15.6,4.7) &\sigma_{\text{GW}}=
\begin{pmatrix}
0.28 & -0.0026 \\
-0.0026 & 0.12
\end{pmatrix}
\qquad\qquad\text{(model 1)}\,,\qquad\qquad\qquad\nonumber\\
\mu_{\text{GW}} &= (-14.9, 4.45) &\sigma_{\text{GW}}=
\begin{pmatrix}
0.059 & -0.0845 \\
-0.0845 & 0.35
\end{pmatrix}
\qquad\qquad\text{(model 2)}\,.\qquad\qquad\qquad
\end{align}
The corresponding $\chi^2$-distribution is defined by
\begin{equation}\label{eq:chimodel}
\chi^2_{\text{model}}= \Delta_{\text{GW}}
\;\sigma^{-1}_{\text{GW}}\; \Delta_{\text{GW}}^T\,\qquad\text{with}\quad \Delta_{\text{GW}}=(\log_{10}A_{\text{GW}},\gamma_{\text{GW}})-\mu_{\text{GW}}\,,
\end{equation}
where $\sigma^{-1}_{\text{GW}}$ stands for the inverse matrix of $\sigma_{\text{GW}}$.

\begin{figure}[h!]
\begin{center}
\includegraphics[width=0.95\textwidth]{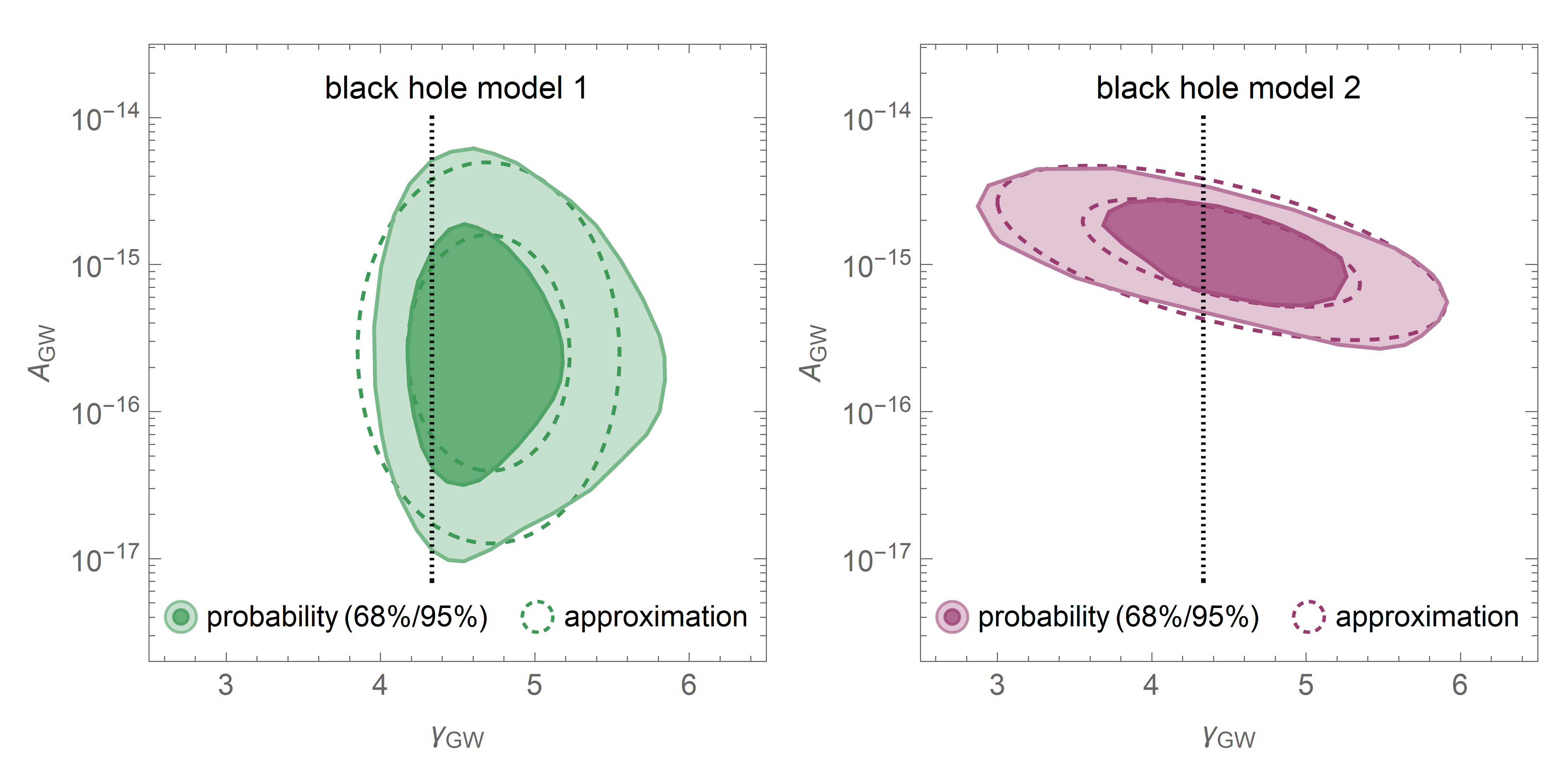}
\end{center}
\vspace{-5mm}
\caption{Probability Regions ($68\%/95\%$) in terms of the GW amplitude and spectral index (defined in Eq.~\eqref{eq:omegahc}) for the two black hole models employed in this work. Also shown are the regions obtained with the Gaussian approximation (described in the text). The black dotted lines represent the prediction $\gamma_{\text{GW}} = 13/3$ of the (old) reference black hole model.}
\label{fig:blackhole}
\end{figure}

The values of $\mu_{\text{GW}}$ and $\sigma_{\text{GW}}$ in Eq.~\eqref{eq:SMBHparameters} for model 1 were provided in~\cite{NANOGrav:2023hvm}, while the values for model 2 have been obtained by us through a fit to the full probability distributions. The probability contours ($68\%/95\%$) in the ($A_{\text{GW}}$,$\gamma_{\text{GW}}$)-plane as well as their Gaussian approximations are depicted in Fig.~\ref{fig:blackhole} for both black hole models.\footnote{In the Gaussian approximation the $68\%$ and $95\%$ probability contours correspond to $\chi_{\text{model}}^2=2.4$ and $\chi_{\text{model}}^2=6.0$ respectively.}

\section{Gravitational Waves from First-Order Phase Transitions}\label{sec:fopt}

While SMBH mergers present a plausible astrophysical explanation of a stochastic GW background, there also exist well-motivated cosmological alternatives. In particular, a FOPT in the early universe has long been considered as a source of detectable gravitational radiation~\cite{Witten:1984rs,Hogan:1986qda}. Because the GW spectrum induced by a FOPT depends on the underlying microphysics we will focus on three main cases in the following: (i) a strong FOPT, (ii) a dark sector FOPT (iii) a thermal FOPT. Before turning to these scenarios in more detail, we will briefly review the calculation of the GW signal associated with a FOPT.

\subsection{Gravitational Wave Spectrum}\label{sec:foptgwspectrum}

A FOPT proceeds by the nucleation, expansion and merger of bubbles of new vacuum within the sea of old vacuum. The expanding and colliding bubbles can be the source of strong gravitational radiation~\cite{Witten:1984rs,Hogan:1986qda}. The latter is generated because the bubble collisions (BC) break spherical symmetry and lead to a non-zero time-dependent tensor anisotropic stress~\cite{Kosowsky:1992rz,Kosowsky:1992vn}. Furthermore, GWs can also be induced by sound waves (SW)~\cite{Hindmarsh:2013xza,Hindmarsh:2015qta,Hindmarsh:2017gnf} and magnetohydrodynamic turbulence~\cite{Kosowsky:2001xp,Dolgov:2002ra,Caprini:2009yp} which form if the expanding bubbles interact with a surrounding radiation plasma. The size of the different contributions depends on the underlying microphysics: BC tend to dominate the gravitation wave production if bubble-plasma interactions are suppressed or absent during the FOPT (for instance if the FOPT occurs in vacuum), while SW and magneto-hydrodynamic turbulence tend to dominate in the opposite regime. In the following we will focus on bubble- and sound-wave-induced GWs because the magnetohydrodynamics contribution is likely subdominant (and difficult to pin down) (see e.g.~\cite{Caprini:2018mtu}).

\paragraph*{Gravitational Waves from Bubble Collisions\\}
The GW spectrum from BC is commonly estimated in the ``envelope approximation'' which places the stress-energy in an infinitesimally thin shell at the bubble wall which immediately disappears upon collision~\cite{Kosowsky:1992rz,Kosowsky:1992vn,Huber:2008hg}. It has, however, been argued that the envelope approximation only applies to FOPTs in the thin-wall regime of vacuum tunneling. This is because in thin-wall collisions the tunneling field becomes trapped temporarily in the old vacuum within the bubble collision region~\cite{Hawking:1982ga,Watkins:1991zt,Falkowski:2012fb} leading to a minimization of the shear stress after collision~\cite{Jinno:2019bxw}. However, in the opposite thick-wall regime, no trapping of the tunneling field occurs and the latter rather undergoes oscillations around the new vacuum within the bubble overlap region. This causes substantial propagation of the shear stress after collision, thus invalidating the assumption of the envelope approximation~\cite{Cutting:2020nla}. Simulations in the thick-wall regime have, for instance, been performed in~\cite{Cutting:2020nla}\footnote{The thick-wall case corresponds to the smallest choice of $\bar{\lambda}$ in~\cite{Cutting:2020nla}.} and predict a GW spectrum significantly distinct from the envelope approximation. 
Such differences -- which mostly affect how quickly the spectrum falls away from the peak -- should not be considered as an uncertainty, rather they reflect the distinct physical realities (thin-wall bubbles vs.\ thick-wall bubbles). In order to capture the relevant cases, we will implement the GW signal from BC separately for the envelope approximation~\cite{Kosowsky:1992rz,Kosowsky:1992vn,Huber:2008hg} and the thick-wall simulation of~\cite{Cutting:2020nla}.

Note, however, that simulations were optimized to predict the GW spectrum around the peak frequency and may not well describe the far-infrared and far-ultraviolet regime. In particular, we need to take into account causality considerations, which impose that the GW spectrum in the far-infrared turns over to a white noise spectrum, i.e. a power-law with index 3~\cite{Caprini:2009fx}. Bearing this in mind we will model the GW spectrum from BC by a broken power law~\cite{Kosowsky:1992rz,Kosowsky:1992vn} of the form,
\begin{equation}\label{eq:gwbubbles}
\Omega_{\text{GW}}^{\text{BC}}h^2 = \Omega_{\text{GW,peak}}^{\text{BC}}h^2\times
\left(\frac{(a+b)\left(\frac{f}{f_{\text{peak}}}\right)^{\frac{a}{c}}}{b+a\left(\frac{f}{f_{\text{peak}}}\right)^{\frac{a+b}{c}}}\right)^c\times 
F_{\text{causal}}(f)\,,
\end{equation}
where $\Omega_{\text{GW,peak}}$, $f_{\text{GW,peak}}$ denotes the peak density and peak frequency in today's universe. The power-law indices $a,b$ and the smoothness of the break $c$ for the thin-wall (=envelope approximation) and thick-wall case are presented in Tab.~\ref{tab:bcparameters}. They were extracted from~\cite{Huber:2008hg,Jinno:2016vai,Lewicki:2020azd}\footnote{In~\cite{Huber:2008hg} a simpler parameterization with $c=1$ has been assumed, for which $a=2.8$ was obtained by a fit procedure. However, we find that the fit to the simulation data presented in~\cite{Huber:2008hg} can be improved for $a=3$, $c=2$. This choice is also consistent with~\cite{Jinno:2016vai,Lewicki:2020azd}.} and~\cite{Cutting:2020nla} respectively. The factor $F_{\text{causal}}(f)$ accounts for the infrared-asymptotic behavior dictated by causality. Since the thin-wall case already features an infrared-index of $a=3$ (cf.~Tab.~\ref{tab:bcparameters}) a non-trivial factor only needs to be considered in the thick-wall case. We, therefore, define (see~\cite{Ellis:2023oxs}),
\begin{equation}\label{eq:Fcausal}
F_{\text{causal}}(f) = \begin{cases} 1 & \;\text{(thin-wall)\,,}\\[1mm]
\left(1+\left(\frac{H_*}{2\pi f_{\text{peak},*}}\right)^{3-a}\right)^{-1} & \;\text{(thick-wall)\,,}
\end{cases}
\end{equation}
where $H_*$ and $f_{\text{peak},*}$ denote the Hubble scale and the peak frequency at the time of the FOPT. The above choice ensures that the GW spectrum follows a white-noise behavior $\Omega_{\text{GW}}^{\text{BC}}h^2\propto f^3$ at scales much larger than all scales in the problem (i.e.\ when the wavelength exceeds the Hubble horizon at the time of the FOPT).

The GW peak energy density and peak frequency can be extracted from the simulations~\cite{Huber:2008hg,Cutting:2020nla}. Including the redshifting between emission and today (assuming a standard post-phase-transition cosmology without entropy production) one obtains,
\begin{align}
\label{eq:Omegabc}
\Omega_{\text{GW,peak}}^{\text{BC}} h^2  &= \left(\frac{H_*}{\beta}\right)^2\,\left(\frac{\kappa_\phi \alpha}{1+\alpha}\right)^2\, \left(\frac{10}{g_{\text{eff}}(T_*)}\right)^{1/3}\times
\begin{cases}
2.7\times 10^{-6} & \text{(thin-wall)\,,}\\[1mm] 9.5\times 10^{-7}& \text{(thick-wall)\,,}
\end{cases}\\
f_{\text{peak}} &= 1.1\times 10^{-7}\:\text{Hz}\;\,\left(\frac{f_{\text{peak,*}}}{H_*}\right)\,\left(\frac{g_{\text{eff}}(T_*)}{10}\right)^{1/6}\left(\frac{T_*}{\text{GeV}}\right)\quad\text{with}\quad f_{\text{peak,*}}=0.2\beta\,,
\label{eq:fpeak}
\end{align}
where $\beta$ is the inverse duration of the FOPT, which is defined as
\begin{equation}
    \beta =-\left.\frac{\dot P}{P}\right|_{t=t_*}\,,
\end{equation}
with $P(t)$ denoting the probability of finding a point in the false vacuum at the time $t$, and $t_*$ the time of the phase transition.\footnote{A discussion of how to derive $\beta$ for time-independent and time-dependent tunneling rates can be found in Sec~2.2 of~\cite{Freese:2022qrl}.}

Furthermore, $\kappa_\phi$ specifies the energy fraction carried by the bubble walls at collision, $T_*$ is the temperature of the radiation bath immediately after the FOPT and $g_{\text{eff}}$ the number of relativistic degrees of freedom (dof). Finally, $\alpha$ stands for the strength of the FOPT which is defined as~\cite{Kamionkowski:1993fg},
\begin{equation}\label{eq:alpha}
\alpha = \frac{\rho_{\text{vac}}}{\rho_{\text{rad}}(T_n)}\,,
\end{equation}
where $\rho_{\text{vac}}$ denotes the vacuum energy released in the FOPT, $T_n$ the temperature right before the transition and $\rho_{\text{rad}}(T_n)$ the corresponding radiation energy density. In the absence of a preexisting plasma, i.e.\ when the FOPT occurs in vacuum, one sets $\kappa_\phi=1$ and $\alpha=\infty$ such that $\kappa_\phi\alpha/(1+\alpha)=1$ in Eq.~\eqref{eq:Omegabc}.

\begin{figure}[htp]
\begin{center}
\includegraphics[width=0.6\textwidth]{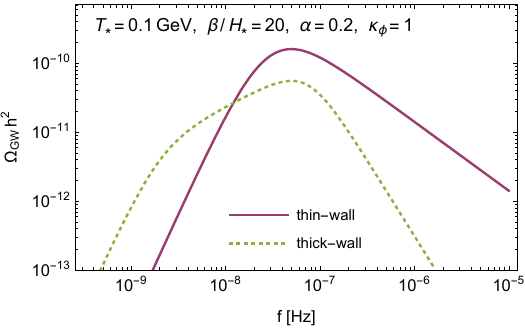}
\end{center}
\vspace{-2mm}
\caption{Comparison of the GW spectra from bubble collisions in the thin-wall and thick-wall regime of vacuum tunneling. The phase transition parameters where chosen as indicated in the figure. In the thick-wall case, there occurs a soft break from the causality tail (with $\Omega_{\text{GW}}\propto f^3$) to a softly rising power-law $\Omega_{\text{GW}}\propto f^{0.7}$ below the peak, and then a second break at the peak frequency, where the spectrum turns into a falling power-law $\Omega_{\text{GW}}\propto f^{-2.2}$. In the thin-wall case, there is is a single break at the peak frequency, where the spectrum changes from $\Omega_{\text{GW}}\propto f^{3}$ to $\Omega_{\text{GW}}\propto f^{-1}$. The amplitude of the spectrum and the location of the spectral breaks can be moved by varying the phase transition parameters.}
\label{fig:gspectra}
\end{figure}

The GW spectra (thin-wall and thick-wall case) from bubble collisions for an example phase transition configuration are depicted in Fig.~\ref{fig:gspectra}. Notice the two spectral breaks in the thick-wall case: one lower-frequency break to account for the transition from the causality tail with $\Omega_{\text{GW}}\propto f^3$ to the below-peak spectrum with $\Omega_{\text{GW}}\propto f^{0.7}$, and one higher-frequency break at the peak, where the spectrum turns into a falling power-law. In contrast, the thin-wall spectrum (=envelope approximation) features only a single break at the peak. Note that the amplitude of the spectrum and the location of the spectral breaks can be moved by varying the phase transition parameters.

\begin{table}[t]
\begin{center}
\begin{tabular}{|cccc|}
\hline
&&&\\[-4mm]
regime
 & $\quad a\quad$ & $\quad b\quad$ & $\quad c \quad$ \\
 \hline
&&&\\[-4mm] 
 thin-wall & $3$ &  $1$ & $2$\\
 thick-wall & $0.7$ &  $2.2$ & $1$\\ \hline
\end{tabular}
\end{center}
\vspace{-0.4cm}
\caption{Parameters entering the GW spectrum from BC in a FOPT (Eq.~\eqref{eq:gwbubbles}) in the thin-wall~\cite{Huber:2008hg,Jinno:2016vai,Lewicki:2020azd} and thick-wall~\cite{Cutting:2020nla} regime of vacuum tunneling. In the thin-wall regime, the envelope approximation is applied.
}
\label{tab:bcparameters}
\end{table}

\paragraph*{Acoustic Gravitational Waves\\}
If the surrounding plasma inflicts substantial  pressure on the bubble walls during their expansion a large fraction of the vacuum energy is transferred to heat and bulk motion of the plasma before the bubbles collide. The bulk motion in the form of sound waves (SW) creates acoustic GWs with a spectrum~\cite{Hindmarsh:2015qta,Caprini:2018mtu},
\begin{equation}\label{eq:gravityspectrumplasma}
\Omega_{\text{GW}}^{\text{SW}} h^2(f)  = \Omega_{\text{GW,peak}}^{\text{SW}} h^2
\left(\frac{f}{f_{\text{peak}}}\right)^3\left(\frac{7}{4+3\left(\tfrac{f}{f_{\text{peak}}}\right)^2}\right)^{\frac{7}{2}}\,,
\end{equation}
where the term $\propto f^3$ ensures the infrared-scaling suggested by causality. Taking into account redshifting (again assuming a standard cosmological evolution) the peak density is determined by~\cite{Hindmarsh:2015qta}
\begin{equation}\label{eq:Omegasound}
\Omega_{\text{GW,peak}}^{\text{SW}} h^2 =
3.5\times 10^{-5}\left(\frac{10}{g_{\text{eff}}(T_*)}\right)^{1/3}\;\left(\frac{H_*}{\beta}\right)\,\left(\frac{\kappa_v \alpha}{1+\alpha}\right)^2\,,
\end{equation}
where we assumed that the bubble walls propagate (approximately) at the speed of light. The efficiency factor $\kappa_v$ denotes the fraction of vacuum energy converted into bulk motion of the plasma~\cite{Espinosa:2010hh},
\begin{equation}\label{eq:kappav}
    \kappa_v \simeq \frac{\alpha}{0.73+0.083\sqrt{\alpha}+\alpha}\,.
\end{equation}
Finally, according to recent simulations~\cite{Hindmarsh:2017gnf}, the peak frequency of the GWs from SW aligns with the peak frequency of the bubble-induced spectrum. Therefore,~Eq.~\eqref{eq:fpeak} also applies to acoustic GWs. 

\subsection{Cases of First-Order Phase Transitions}\label{sec:cases}

As described in the previous section the GW spectrum from a FOPT in the early Universe depends on a number of parameters. Previous works have analyzed the phase-transition hypothesis for the observed stochastic GW background by scanning the entire available parameter space~\cite{NANOGrav:2021flc,NANOGrav:2023hvm,EPTA:2023xxk,Ellis:2023oxs,Figueroa:2023zhu}. While such a general approach provides valuable insights into the general consistency of PTA data with a FOPT, it is less efficient in constraining the underlying physics responsible for the phase transition. In this work we follow a different strategy, where we consider three well-motivated subcases of FOPTs: a strong FOPT, a dark-sector FOPT and a thermal FOPT. Each subcase yields a constrained parameter space with only two free parameters. In this way we do not only test the black-hole merger hypothesis against the phase-transition hypothesis, but we additionally gain insights into the type of phase-transition preferred by the data (if any). In the following, we will describe the three subcases in more detail and present the resulting gravitational radiation signal. For convenience we also provide Tab.~\ref{tab:gwcalc} which allows for a quick look-up of the GW spectrum of the phase-transition scenarios considered. In all cases we will impose that the phase transition completes before BBN -- which translates to $T_*>1.8\:\text{MeV}$~\cite{Hannestad:2004px,Hasegawa:2019jsa} -- such that the light element abundances are not spoiled. Furthermore, successful bubble percolation during the FOPT imposes $\beta/H_*>3$. Otherwise the phase transitions runs into the infamous empty-universe problem (which has first been pointed out in the context of old inflation)~\cite{Guth:1982pn}.

\begin{table}[htp]
\begin{center}
\begin{tabular}{|cccccccccc|}
\hline
&&&&&&&&&\\[-4mm]
  $\!$FOPT type$\!$  & $\!\!$GW source$\!\!$ & $\Omega_{\text{GW}}$ & $\! a,b,c\!$ & $\Omega_{\text{GW,peak}}$ & $f_{\text{peak}}$  & $\alpha$& $\!\kappa_\phi\!$ & $\kappa_v$ & $\beta/H_*$ \\
 \hline
&&&&&&&&&\\[-4mm]
strong & BC & Eq.$\,$\eqref{eq:gwbubbles},\eqref{eq:Fcausal} & $\!$Tab.$\,$\ref{tab:bcparameters}$\!$&Eq.$\,$\eqref{eq:Omegabc}& Eq.$\,$\eqref{eq:fpeak}& $\infty$ & $1$ & $0$ & free\\[1mm]
dark & BC & Eq.$\,$\eqref{eq:gwbubbles},\eqref{eq:Fcausal} & $\!$Tab.$\,$\ref{tab:bcparameters}$\!$ &Eq.$\,$\eqref{eq:Omegabc}& Eq.$\,$\eqref{eq:fpeak} & $\!$free$\!$& $1$ & $0$& Eq.$\,$\eqref{eq:fit}\\[1mm]
thermal & SW & Eq.$\,$\eqref{eq:gravityspectrumplasma} & $-$ & Eq.$\,$\eqref{eq:Omegasound}& Eq.$\,$\eqref{eq:fpeak} & $\!$free$\!$& $0$ & Eq.$\,$\eqref{eq:kappav}& free\\[-4mm] 
 &&&&&&&&& \\ \hline
\end{tabular}
\end{center}
\vspace{-0.4cm}
\caption{Calculation of the GW spectrum for the three cases of FOPTs described in Sec.~\ref{sec:cases}. Here BC refers to Bubble Collisions and SW to sound waves as the predominant origin of the GWs in different cases; $\alpha$ as defined in Eq. \eqref{eq:alpha} determines the strength of the phase transition; $\kappa_\phi$ and $\kappa_v$ are the fractions of the vacuum energy carried by the bubble walls and  converted into bulk motion of the
plasma respectively; and $\beta$ is the inverse duration of the phase transition.}
\label{tab:gwcalc}
\end{table}

\subsubsection*{Strong Phase Transition}

A strong FOPT occurs in the absence of a preexisting radiation plasma (=phase transition in vacuum), or if the released vacuum energy strongly dominates over the thermal energy of the preexisting plasma. Well-motivated examples include a FOPT at the end of inflation or during the graceful exit from inflation -- as predicted in double-field inflation~\cite{Adams:1990ds,Linde:1990gz} and in chain inflation~\cite{Freese:2004vs,Freese:2005kt,Ashoorioon:2008pj}. Alternatively, a strong FOPT may take place in the post-inflationary era after an initial stage of radiation-domination. In fact, there exist attractive cosmological scenarios in which the early universe encountered a second period of vacuum domination (lasting for e.g.\ $1-10$~e-folds) which is terminated by a strong FOPT. The latter reheats the Universe a second time and plays a major role in shaping the Universe we observe today, producing the dark matter~\cite{Chung:1998ua,Kolb:1998ki,Petraki:2011mv,Falkowski:2012fb,Azatov:2021ifm,An:2022toi,Freese:2023fcr}, triggering baryogenesis (see e.g.~\cite{Konstandin:2011dr,Katz:2016adq,Athron:2022mmm}) and/or diluting unwanted relics from the inflationary epoch~\cite{Lyth:1995ka,Barreiro:1996dx} (like moduli fields or gravitinos predicted in supersymmetric cosmologies). In order to realize a second vacuum-dominated stage, the evolution of the thermal potential must give rise to a regime of strong supercooling. While supercooling does not occur during the phase transitions within the Standard Model of particle physics, simple and well-motivated gauge extensions of the Standard Model easily accommodate a supercooling phase followed by a strong FOPT (see e.g.~\cite{Jaeckel:2016jlh,Jinno:2016knw,Addazi:2017gpt,Hashino:2018zsi,Croon:2018erz,Marzo:2018nov,Breitbach:2018ddu,Baratella:2018pxi,Azatov:2019png,Lewicki:2020azd}). 

Due to the strong subdominance (or absence) of a preexisting plasma the bubbles formed during a strong FOPT tend to reach a ``runaway regime'', in which they expand practically unhindered and all the energy of the phase transition is released by the BC. Hence, the GW spectrum induced by a strong FOPT is determined by Eq.~\eqref{eq:gwbubbles}. The peak density is derived from Eq.~\eqref{eq:Omegabc} by taking the limit $\kappa_\phi=1$ and $\alpha\rightarrow \infty$ (in which the released vacuum energy dominates over the preexisting thermal energy). For the two subcases -- tunneling in the thin-wall regime and in the thick-wall regime -- the GW spectrum of a strong FOPT is fully determined by $\beta/H_*$ and $T_*$.

\subsubsection*{Dark-Sector Phase Transition}

Dark sectors are defined as a collection of particles not subject to the gauge forces of the Standard Model. Such particles may interact with the visible matter either solely through gravity, or through small portal couplings~\cite{Holdom:1985ag,Patt:2006fw}. The existence of a dark sector is motivated by its ability to overcome several gaps in the Standard Model of particle physics -- for instance providing explanations for neutrino masses, the baryon asymmetry or the observed dark matter (see e.g.~\cite{Alexander:2016aln}). Furthermore, dark sectors exhibit a strong theoretical motivation from ultraviolet completions of the Standard Model like string theory, in which the presence of dark sectors is required to ensure mathematical consistency (see e.g.~\cite{Acharya:2016fge}). FOPTs in a dark sector have been the subject of active research (see e.g.~\cite{Schwaller:2015tja,Jaeckel:2016jlh,Croon:2018erz,Bai:2018dxf,Nakai:2020oit,Addazi:2020zcj,Ratzinger:2020koh,Borah:2021ocu,Lewicki:2021xku,Bringmann:2023opz,Fujikura:2023lkn,Addazi:2023jvg}). For instance, a ``Dark Big Bang'' FOPT
could generate the dark matter and dark radiation -- analogous to the production of the visible matter and radiation in the Hot Big Bang~\cite{Freese:2023fcr}.

We will assume the absence (or strong subdominance) of a dark sector radiation bath prior to the phase transition -- otherwise we would essentially arrive at the scenario of a thermal FOPT (to be discussed in the next section). With only visible radiation around, the vacuum bubbles of the phase transition do typically not lose substantial energy during their expansion. This holds even in a radiation-dominated\footnote{By radiation domination we mean domination by the visible sector radiation bath.} universe because of the suppressed coupling of the (dark sector) tunneling field to Standard Model particles. The GW spectrum of a dark FOPT is, hence, determined by Eq.~\eqref{eq:gwbubbles} and Eq.~\eqref{eq:Omegabc} with $\kappa_\phi=1$. In the assumed absence of dark sector thermal effects (prior to the phase transition), the vacuum decay rate (typically) becomes time-independent, and the duration of the phase transition is controlled by the expansion rate\footnote{In the literature on FOPTs the duration of the phase-transition it often approximated by $\beta=\dot{\Gamma}/\Gamma$ with $\Gamma$ denoting the tunneling rate. However, this approximation breaks down in the absence of thermal effects because the tunneling rate typically becomes time-independent in this limit such that\ $\dot{\Gamma}=0$. As a consequence, the expansion rate rather than $\dot{\Gamma}$ controls the duration of the FOPT (see discussion in~\cite{Freese:2022qrl}).} -- which is set by $\alpha$. The precise relation between $\beta$ and $\alpha$ for a FOPT with a time-independent vacuum decay rate can be extracted from Fig.~1 of~\cite{Freese:2022qrl}. By performing a numerical fit to this figure, we find the following approximate relation,
\begin{equation}\label{eq:fit}
\frac{\beta}{H_*}\simeq \frac{8-0.64\,\log(\alpha+1)}{1+0.32\,\log(\alpha+1)}\,.
\end{equation}
With $\beta/H_*$ fixed by the equation above, the GW spectrum of a dark FOPT depends on the two parameters $\alpha$ and $T_*$. Again, we will consider the two subcases of tunneling in the thick-wall and thin-wall regime.

Let us point out that successful bubble percolation during the FOPT imposes
\begin{equation}
 \frac{\beta}{H_*}>3 \quad\Longleftrightarrow \quad\alpha <21\,.
\end{equation}
Otherwise the phase transitions runs into the infamous empty-universe problem (originally pointed out in the context of old inflation)~\cite{Guth:1982pn}. The fit in Eq.~\eqref{eq:fit} is valid in the entire regime in which the bubble percolation condition is satisfied.

Finally, we note that a phase transition in a fully decoupled dark sector is strongly constrained by the cosmic microwave background due to the additional dark sector radiation energy density present in the Universe after the phase transition. Depending slightly on the composition of the dark sector a constraint in the range $\alpha\lesssim 0.1$ would apply~\cite{Planck:2018vyg,Nakai:2020oit,Freese:2023fcr}. 
However, this bound is evaded if small portal couplings between the dark and the visible sector exist, which induce the subsequent decay of the dark radiation states into visible radiation after the FOPT (see e.g.~\cite{Bringmann:2023opz}). Allowing for the presence of such couplings, we will not impose the constraint on $\alpha$ in the following. We mention that if dark-radiation decay occurs, the phase transition should typically complete before BBN ($T_*>1.8\:\text{MeV}$) in order to avoid that the energy injection into the visible sector spoils the light element abundances. We will find, however, that the dark sector FOPTs best-fitting the PTA signals features a larger $T_*$ such the BBN constraint effectively plays no role.

\subsubsection*{Thermal Phase Transition}

So far, we discussed FOPTs in vacuum, within a highly subdominant radiation environment and in a secluded sector -- cases which allow for the vacuum bubbles to propagate (almost) freely and produce most of the gravitational radiation by their collision. A contrary situation occurs for a FOPT in the visible sector during a radiation-dominated period. The properties of such a thermal phase transition are controlled by the interactions of the tunneling scalar field with the thermal plasma. Once the bubbles of true vacuum form and expand, these same interactions inflict friction on the bubble walls which -- after initial
acceleration -- reach a constant velocity $v_w$ (since the size of $v_w$ does not considerably affect our fits to the PTA signals\footnote{The velocity $v_w$ mostly affects the normalization and peak frequency of the GW spectrum, but not the spectral shape. In our analysis, where we investigate the general consistency of the observed PTA signal with a FOPT, we will take the normalization and peak frequency as free parameters. Therefore, a different choice of $v_w$ would typically not affect our fits, and we can fix $v_w\simeq c$ without loosing generality.}, we will simply set $v_w\simeq c$). While the bubbles expand into the plasma, they induce compression waves -- sound waves -- which act as a strong source of gravitational radiation. In fact, simulations have revealed SW as the most relevant contribution to the GW spectrum of phase transitions which are not very strongly first-order and happen in a thermal environment~\cite{Hindmarsh:2013xza,Hindmarsh:2015qta}. We can thus approximate the spectrum of a thermal phase transition by Eq.~\eqref{eq:gravityspectrumplasma} with the peak density from Eq.~\eqref{eq:Omegasound} and the efficiency factor (which accounts for the fact that some of the phase-transition energy is lost to heat not contributing to the GW production) from Eq.~\eqref{eq:kappav}.

\section{Gravitational Wave Signal at Pulsar Timing Arrays}\label{sec:signal}

The NANOGrav~\cite{NANOGrav:2020bcs,NANOGrav:2023gor}, PPTA~\cite{Goncharov:2021oub,Reardon:2023gzh}, EPTA~\cite{Chen:2021rqp,EPTA:2023fyk} and CPTA~\cite{Xu:2023wog} collaboration have detected a low-frequency stochastic process which affects pulsar timing residuals. There is, furthermore, evidence that the signal exhibits the quadrupolar Hellings-Downs correlations~\cite{Hellings:1983fr} associated with GW sources~\cite{NANOGrav:2023gor,NANOGrav:2023hvm,EPTA:2023fyk,EPTA:2023xxk,Reardon:2023gzh}. The PTA observations, hence, likely amount to the discovery of a stochastic GW background in the nHz-frequency regime. 

\subsection{Data Extraction and Fit Method}
In this analysis we will consider the following data sets
\begin{itemize}
\item NANOGrav's 15-yr data set~\cite{NANOGrav:2023gor,NANOGrav:2023hvm}
\item PPTA's 18-yr data set~\cite{Reardon:2023gzh}
\item EPTA's newly released data set~\cite{EPTA:2023fyk,EPTA:2023xxk} of which two sub-versions exist which we denote as EPTA$_1$ and EPTA$_2$ in the following. EPTA$_1$ includes only the last 10.3-yr of data based on modern observing systems, while EPTA$_2$ is the full 24.7-yr data set (including lower quality early data).
\end{itemize}
We refrain from including the CPTA data set~\cite{Xu:2023wog} due to its significantly larger error bars.

\begin{figure}[h!]
\begin{center}
\includegraphics[width=0.95\textwidth]{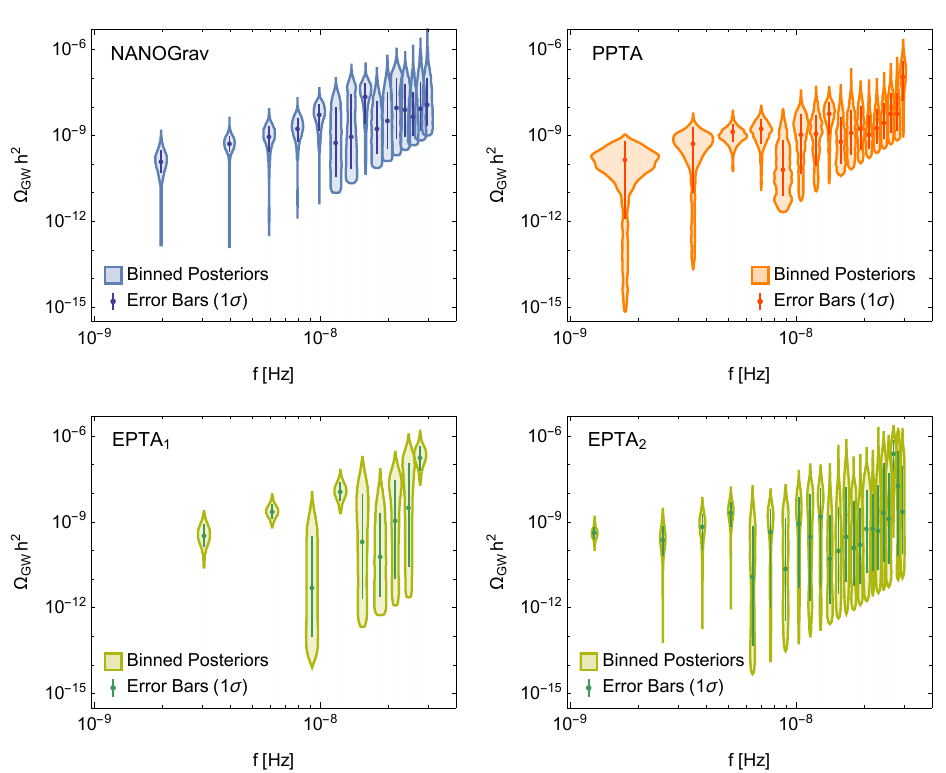}
\end{center}
\vspace{-2mm}
\caption{Spectrum of the GW signal observed by the different PTA experiments. Shown are the published posterior distributions for a free-spectrum GW process (violins)~\cite{NANOGrav:2023gor,NANOGrav:2023hvm,Reardon:2023gzh,EPTA:2023fyk,EPTA:2023xxk} as well as the medians and 1$\sigma$ error bars constructed in this work.}
\label{fig:dataextraction}
\end{figure}

In order to extract the observed GW spectra $\Omega_{\text{GW}}(f)$ we digitized the published posterior distributions for a free-spectrum GW process~\cite{NANOGrav:2023hvm,Reardon:2023gzh,EPTA:2023fyk,EPTA:2023xxk} which corresponds to the independent measurement of common power at each frequency bin (blue, orange and green violins in Fig.~\ref{fig:dataextraction}). For eliminating high-frequency noise we only included the frequency bins with $f<3\times 10^{-8}\:\text{Hz}\simeq 1\:\text{yr}^{-1}$ in our analysis following~\cite{NANOGrav:2023gor,NANOGrav:2023hvm,EPTA:2023fyk,EPTA:2023xxk,Reardon:2023gzh}. From the posterior distributions we determined the median $\Omega_{\text{GW}}(f_i)$ as well as the upper and lower 1-sigma error of the free-spectrum in each frequency bin $f_i$. The so-obtained error bars are also depicted in Fig~\ref{fig:dataextraction}. In total, we have 15 data points for NANOGrav, 17 for PPTA, 9 for EPTA$_1$ and 22 for EPTA$_2$.

In the following we will interpret the error bars (shown in Fig~\ref{fig:dataextraction}) as statistically independent measurements (and uncertainties) of the GW spectrum. The posterior distributions will be approximated as Gaussian in $\log_{10}\Omega_{\text{GW}}$ which turns out to be a reasonable approximation since significant non-Gaussianity only occurs in those frequency bins with the highest uncertainty, i.e.\ which carry the lowest statistical weight. The Gaussian approximation allows us to employ a standard $\chi^2$ goodness-of-fit metric. For testing the compatibility of a black-hole or phase-transition model with the data of a given experiment (NANOGrav, PPTA or EPTA), we, hence, define,
\begin{equation}\label{eq:chi2}
\chi^2_{\text{experiment}}=\sum\limits_i \left(\frac{\log_{10}\Omega_{\text{GW},i}^{\text{model}}-\log_{10}\Omega_{\text{GW},i}^{\text{data}}}{\Delta\log_{10}\Omega_{\text{GW},i}^\text{data}}\right)^2\,,
\end{equation}
where the sum runs over the bins of the respective experiment (NANOGrav, PPTA, EPTA$_1$ or EPTA$_2$). In this equation $\Omega_{\text{GW},i}^{\text{model}}$ and $\Omega_{\text{GW},i}^{\text{data}}$ stand for the model prediction and the measured spectrum (=median of the free spectrum process) in the $i$th bin, while 
$\Delta\log_{10}\Omega_{\text{GW},i}^\text{data}$ stands for the corresponding 1$\sigma$-error. Due to the asymmetry of the error bars we enter the upper or lower error in Eq.~\eqref{eq:chi2} depending on whether the model prediction lies above or below the measured value.

When analyzing the experimental data sets (after the frequency-cut) we still found evidence for a white-noise remnant in EPTA$_2$ (consistent with the discussion in~\cite{EPTA:2023fyk}). The noise contamination was identified following~\cite{NANOGrav:2020bcs} by performing a power-law vs.\ broken power-law fit to the measured spectrum, where a white-noise scaling $\Omega_{\text{GW,peak}}\propto f^3$ was assumed in the broken power-law fit above the break. Employing the Akaike Information Criterion (AIC)~\cite{Akaike}, we found a clear preference for the fit including the white-noise tail ($\Delta\text{AIC}=4.4$) in EPTA$_2$. At the same time neither of the other data sets (NANOGrav, PPTA, EPTA$_1$) exhibit any statistical preference for a white-noise tail (according to AIC) -- suggesting that this is indeed a noise contamination of EPTA$_2$ and not a GW signal which accidentally scales like white noise. The noise interpretation is reaffirmed by the fact that quadrupolar Hellings-Downs correlations have been established in the EPTA$_1$ but not in the EPTA$_2$ data set -- even though the latter corresponds to a longer observation time~\cite{EPTA:2023fyk}. For the above arguments we will employ EPTA$_1$ (and not EPTA$_2$) in our interpretation of the GW signals. The total goodness-of-fit of a given model with respect to the joined PTA data is, hence, defined as
\begin{equation}\label{eq:chitot}
\chi_\text{PTA}^2 = \chi_\text{NANOGrav}^2 + \chi_\text{PPTA}^2 + \chi_{\text{EPTA}_1}^2\,,
\end{equation}
with the $\chi^2$ for the individual experiments defined in Eq.~\eqref{eq:chi2}.

\subsection{Validation of the Data-Extraction Procedure}

In order to validate our data-extraction method (including the Gaussian error approximation), we will now derive frequentist confidence intervals for a power-law fit to the PTA data. The power-law parameterization is taken from from Eq.~\eqref{eq:omegahc}, where the amplitude $A_{\text{GW}}$ and power-law index $\gamma_{\text{GW}}$ are free parameters.\footnote{The power-law model employs the same parameterization as the SMBH models. However, in the power-law model the amplitude and power-law index are unconstrained in contrast to the SMBH models.} For each experiment we determine the best-fit point and the corresponding $\chi^2_{\text{best}}$. Since the fit involves two free parameters, the $2\sigma$-contour is defined by $\Delta\chi^2=\chi^2-\chi_{\text{best}}^2=6.2$. 

\begin{figure}[htp]
\begin{center}
\includegraphics[width=0.98\textwidth]{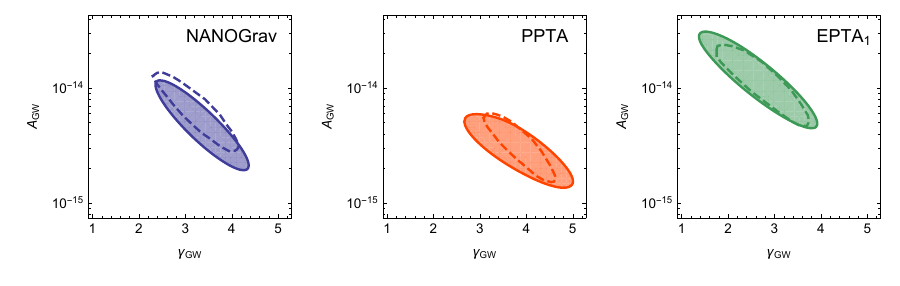}
\end{center}
\vspace{-2mm}
\caption{Confidence intervals (2$\sigma$) obtained in our power-law fit to the NANOGrav, PPTA and EPTA$_1$ data (colored regions) for amplitude $A_{GW}$ and power law spectral index $\gamma_{GW}$ as defined in Eq. \eqref{eq:omegahc} for merging SMBHs. Also shown are the 2$\sigma$-contours of the Bayesian posterior distributions published by the NANOGrav, PPTA and EPTA collaborations (dashed lines).}
\label{fig:validation}
\end{figure}

In Fig.~\ref{fig:validation} we present the $2\sigma$-confidence intervals obtained for a power-law fit to the NANOGrav, PPTA and EPTA$_1$ data. These can directly be compared to the 2$\sigma$-contours for the power-law model published by the experimental collaborations~\cite{NANOGrav:2023gor,EPTA:2023fyk,Reardon:2023gzh} (also shown in Fig.~\ref{fig:validation}). It can seen that our fit results are in reasonable agreement with the published ones. Minor differences arise due to the different statistical interpretation -- while we derived frequentist $2\sigma$-confidence intervals, the experimental groups have provided the 2$\sigma$-contours of the Bayesian posterior distributions. Furthermore, our simplified Gaussian treatment of error bars may play a small role. However, the overall good agreement validates our data extraction and fit method which we can, hence, confidently apply to the SMBH- and FOPT-scenarios which are the focus of this work.

\section{Hypothesis Test: First-Order Phase Transition vs.\ Supermassive Black Hole Mergers}\label{sec:test}

We now turn to our main analysis, where we test the SMBH- against the FOPT-interpretation of the GW signal observed at the PTA experiment. Compared to previous work on the topic, there are three main differences: (1) we will perform a joined fit to the data of all three major PTA experiments instead of a single one, (2) we will consider physically-motivated subcases of FOPTs instead of scanning the entire FOPT parameter space (see Sec.~\ref{sec:cases}), and (3) our analysis relies on frequentist hypothesis testing in contrast to the Bayesian approach of~\cite{NANOGrav:2023hvm,EPTA:2023xxk,Ellis:2023oxs,Figueroa:2023zhu}. 
Since FOPT scenarios are essentially unconstrained experimentally, Bayesian analyses of the PTA data in terms of FOPTs have been found to be highly sensitive to prior choices (see e.g.\ discussion in~\cite{Bringmann:2023opz}) -- making their interpretation challenging. We can evade this issue by employing frequentist statistics.

\subsection{Testing the Supermassive Black Hole Hypothesis}

In the first step, we will test the overall consistency of the SMBH hypothesis with the observed signal by means of a goodness-of-fit test. For this purpose we will employ the joined PTA data set (including NANOGrav, PPTA and EPTA$_1$) and identify the best-fit points of our two SMBH merger models -- model~1 which assumes circular, purely gravitational-wave-driven binary evolution and model~2 which incorporates eccentricities and environmental couplings (see Sec.~\ref{sec:gwbhm}).

The GW spectrum is parameterized by Eq.~\eqref{eq:omegahc}. In order to assess the goodness-of-fit to the PTA data we consider the $\chi_{\text{PTA}}^2$-metric defined in Eq.~\eqref{eq:chitot} and Eq.~\eqref{eq:chi2}. Additionally, we need to take into account the theoretical constraints on the model parameters as stated in Eq.~\eqref{eq:SMBHparameters} (and illustrated in Fig.~\ref{fig:blackhole}). This is done through the separate metric $\chi_\text{model}^2$ defined in Eq.~\eqref{eq:chimodel}. Hence, for determining the best-fit points in the two SMBH models we minimize 
\begin{equation}
\chi_\text{tot}^2=\chi_\text{PTA}^2 + \chi_\text{model}^2\,.
\end{equation}
The resulting best-fit points and their $\chi^2$-values are presented in Tab.~\ref{tab:SMBHbest}. For convenience we also provide $\chi_\text{tot}^2/\text{dof}$, where the number of dof we have to consider is 41.\footnote{The number of dof corresponds to the sample size (41 PTA bins + 2 constraints on parameters) minus the number of parameters (2). Hence we obtain 41+2-2=41 for the number of dof.} The corresponding best-fit GW spectra are depicted in the left panel of Fig.~\ref{fig:spectra}.

\begin{table}[hpt]
\begin{center}
\begin{tabular}{|ccccccc|}
\hline
&&&&&&\\[-4mm]
model  &  $A_{\text{GW}}$ & $\gamma_{\text{GW}}$ & $\chi_\text{PTA}^2$ & $\chi_\text{model}^2$& $\chi_\text{tot}^2$& $\chi_\text{tot}^2/\text{dof}$\\
 \hline
&&&&&&\\[-4mm] 
 1& $3.4\times 10^{-15}$ & $3.93$ & $34.7$ & $9.4$ & $44.1$ & $1.08$\\
 2& $4.1\times 10^{-15}$ & $3.64$ & $31.7$ & $4.8$ & $36.5$ & $0.89$\\ \hline
\end{tabular}
\end{center}
\vspace{-0.4cm}
\caption{Best-fit amplitude $A_{\text{GW}}$ and power-law index $\gamma_{\text{GW}}$ of the two SMBH models considered in this work. The corresponding $\chi^2$-values for the fit to the PTA data and the SMBH model constraints as well as the total $\chi^2$ are also provided.}
\label{tab:SMBHbest}
\end{table}

\begin{figure}[htp]
\begin{center}
\includegraphics[width=0.37\textwidth]{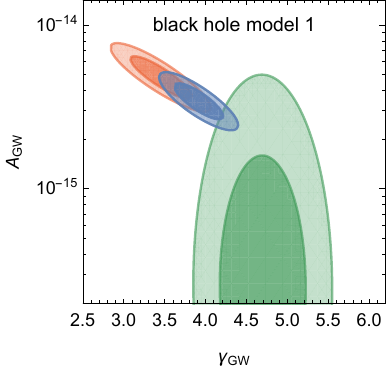}\hspace{1mm}
\includegraphics[width=0.37\textwidth]{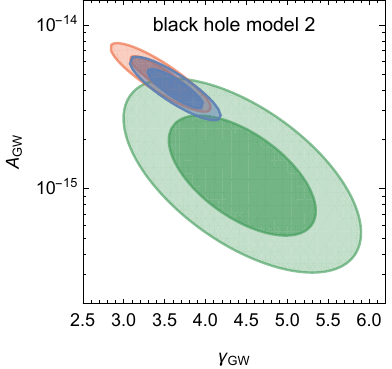}\hspace{1mm}
\includegraphics[trim= 0 -77 0 0,width=0.23\textwidth,clip]{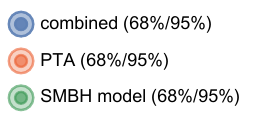}
\end{center}
\vspace{-2mm}
\caption{Confidence regions (68\%/95\%) in the two SMBH models considered in this work for amplitude $A_{GW}$ and power law spectral index $\gamma_{GW}$ as defined in Eq. \eqref{eq:omegahc}. The red contours are obtained by a fit to the PTA data, while the green contours indicate the black hole model constraints. The joined fit incorporating PTA data and model constraints gives rise to the blue contours.}
\label{fig:bhfit}
\end{figure}

\begin{figure}[htp]
\begin{center}
\includegraphics[width=\textwidth]{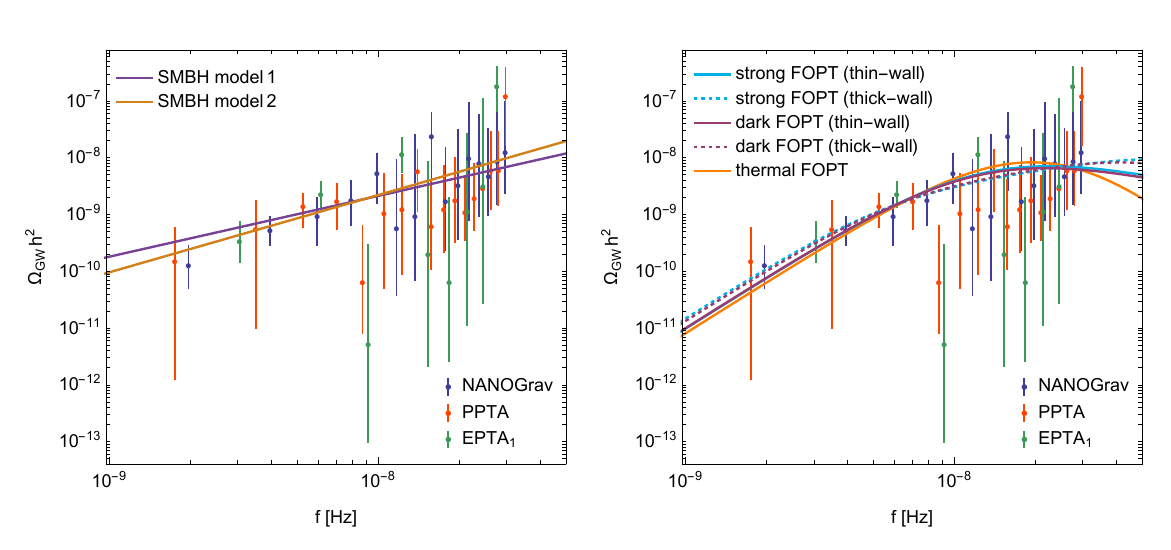}
\end{center}
\vspace{-2mm}
\caption{Best-fit GW spectra in the SMBH merger models (left panel) and in the FOPT scenarios (right panel) considered in this work. Notice that in the right panel the purple lines are almost on top of the cyan lines signaling that the best-fit spectra of the strong and the dark FOPT are very similar.
}
\label{fig:spectra}
\end{figure}

We observe that both black hole models feature $\chi_\text{tot}^2/\text{dof}\sim 1$. From a pure goodness-of-fit standpoint they are, hence, both compatible with the observed GW signal (within $1\sigma$). However, if we compare the $\chi_\text{tot}^2$-values it becomes evident that model~2 provides a significantly better fit than model~1. This can be understood by looking separately at the confidence regions associated with the PTA signal and with the black hole model constraints in Fig.~\ref{fig:bhfit}. The PTA signal (red region) prefers a larger $A_{\text{GW}}$ and smaller $\gamma_{\text{GW}}$ than what is favored by the black hole population underlying model~1 (green region) -- in agreement with the findings of~\cite{NANOGrav:2023hvm}

A better agreement between PTA data and black hole model constraints is obtained in model 2 (although the PTA-observed amplitude is still a bit on the high side). This is due to the inclusion of environmental couplings and eccentricities affecting the binary orbits in model~2 which lead to a broader probability distribution in terms of $\gamma_{\text{GW}}$ -- in contrast to model~1 where circular SMBH orbits and purely gravitational-wave driven evolution imply a spectral index close to the canonical prediction of $\gamma_{\text{GW}}=13/3$.

While the fit results indicate a preference for model~2 we remind the reader that model~2 has been constructed after the release of the PTA data. It amounts to an attempt to include effects in the SMBH dynamics which improve the consistency of the PTA signal with the black-hole hypothesis. Whether these effects (eccentricities and environmental couplings) play any role in real SMBH binary systems is unknown. Once we do not impose the SMBH hypothesis for the GW signal, there is limited motivation for the large eccentricities included in model~2 and the SMBH-induced GW signal may be closer to model~1 -- or even completely negligible if no solution to the final-parsec problem is found. When testing the SMBH hypothesis for the PTA signal against the FOPT hypothesis, we will thus not disregard model~1.

\subsection{Testing the First-Order Phase Transition Hypothesis}

We now turn to the FOPT hypothesis and again perform a goodness-of-fit test on the joined PTA data. For this purpose we consider the three FOPT models of Sec.~\ref{sec:cases} which consist in a strong, a dark and a thermal phase transition. For the strong and the dark FOPT there exist two subcases depending on whether the phase transition occurs in the thin-wall or the thick-wall regime of vacuum tunneling (see Sec.~\ref{sec:foptgwspectrum}). The GW spectrum for each type of phase transition is calculated according to Tab.~\ref{tab:gwcalc}.

Each of the FOPT scenarios exhibits two free parameters which are $T_*$, $\beta/H_*$ for the strong phase transition and $T_*$, $\alpha$ for the dark phase transition.\footnote{As a reminder, $\beta$, $T_*$, and $\alpha$ are the inverse duration, temperature, and strength of the FOPT.  Further, remember that $\beta/H_*$ is fixed in terms of $\alpha$ by Eq.~\eqref{eq:fit} for a dark FOPT.} In the case of a thermal phase transition -- even though there are in principle three free parameters $T_*$, $\beta/H_*$ and $\alpha$ -- the GW spectrum effectively only depends on two parameter combinations (due to a degeneracy). When deriving the goodness-of-fit we can thus fix one parameter without loss of generality which we pick to be $\beta/H_*=10$ (which we do only for the thermal phase transition). Later, when we analyze the preferred parameter space for the thermal phase transition, we will also consider different values of $\beta/H_*$.

In order to assess the goodness-of-fit to the PTA data we again consider the $\chi_{\text{PTA}}^2$-metric defined in Eq.~\eqref{eq:chitot} and Eq.~\eqref{eq:chi2}. In the FOPT-case the model parameters can vary freely within the regime $T_*>1.8\:\text{MeV}$, $\beta/H_*>3$ allowed by BBN and by the percolation condition. Therefore, for finding the best-fit points, we simply need to minimize $\chi_{\text{PTA}}^2$. 

The best-fit points for the strong, dark and thermal phase transition and the corresponding $\chi_{\text{PTA}}^2$ and $\chi_{\text{PTA}}^2/\text{dof}$ values are presented in Tab.~\ref{tab:FOPTbest}. Notice that since the model parameters can vary freely, the FOPT fit has 39 dof (in contrast to the SMBH fit with 41 dof).\footnote{The number of dof is given by the sample size minus the number of parameters. In the FOPT case the sample size is 41 corresponding to the number of PTA bins. In contrast, in the SMBH case the sample size is 43 due to the two additional boundary conditions (on the black hole model parameters).} The corresponding best-fit GW spectra are depicted in the right panel of Fig.~\ref{fig:spectra}.

\begin{table}[hpt]
\begin{center}
\begin{tabular}{|ccccccc|}
\hline
&&&&&&\\[-4mm]
FOPT type  &  bubble wall &$T_*$~[GeV] & $\alpha$ & $\beta/H_*$& $\chi_\text{PTA}^2$& $\chi_\text{PTA}^2/\text{dof}$\\
 \hline
&&&&&&\\[-4mm] 
 strong & thin& $0.051$ &  $\infty$ & $18.4$ &  $33.9$ & $0.87$\\
 strong &thick& $0.23$ &  $\infty$ & $7.8$ & $31.2$ & $0.80$\\
 dark & thin& $0.13$ &  $0.59$ & $6.8$ & $33.9$ & $0.87$\\
 dark & thick& $0.25$ &  $2.0$ & $5.4$ & $31.4$ & $0.81$\\
 thermal& $-$ & $0.079$ &  $0.5$ & $10$~(fixed) & $36.8$ & $0.94$\\ \hline
\end{tabular}
\end{center}
\vspace{-0.4cm}
\caption{Best-fit parameters and corresponding $\chi^2$-values for a strong, dark and thermal FOPT derived from the fit to the PTA data. For the strong and the dark phase transition the cases of vacuum tunneling in thin- and thick-wall regime are considered separately (see text). For the thermal phase transition, the fit is independent of the wall thickness (because the GW signal is produced by sound waves rather than by bubble collisions).}
\label{tab:FOPTbest}
\end{table}

All considered FOPT scenarios feature $\chi_\text{PTA}^2/\text{dof}\sim 1$. From a goodness-of-fit standpoint all of them are, hence, compatible with the observed GW signal (within $1\sigma$). Among the considered cases a slight preference (lower $\chi_\text{PTA}^2$) is observed for the strong and the dark phase transition within the thick-wall regime of vacuum tunneling. This is because the PTA experiments (slightly) prefer a transition from a steeply rising power-law to a softly rising power-law within their frequency band, as is realized in the best-fit thick-wall spectra of Fig.~\ref{fig:spectra}. The transition from a rising to a falling power-law (as for the best-fit thermal FOPT spectrum of Fig.~\ref{fig:spectra}) provides a slightly worse fit.

\begin{figure}[htp]
\begin{center}
\includegraphics[width=\textwidth]{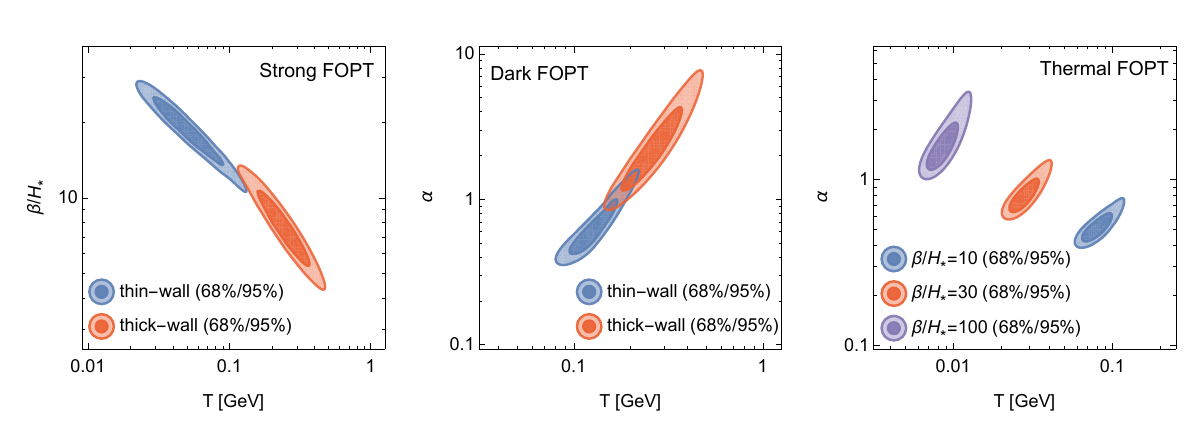}
\end{center}
\vspace{-2mm}
\caption{Confidence regions (68\%/95\%) for the strong, dark and thermal FOPTs obtained by the fit to the PTA signal. For the strong and the dark phase transitions (left and middle panels) we show the confidence regions separately for the thin-wall and thick-wall regime of vacuum tunneling. For the thermal phase transition (right panel) we consider several fixed values of $\beta/H_*$ as indicated in the plot legend.}
\label{fig:FOPTregions}
\end{figure}

In Fig.~\ref{fig:FOPTregions} we present the parameter spaces of the strong, dark and thermal phase transition preferred by the PTA signal. A generic finding within all subcases is that only a strong or moderately strong ($\alpha\gtrsim 0.3$) and relatively slow ($\beta/H_*\lesssim 100$) FOPT is compatible with the observed GW signal.\footnote{In the case of a thermal FOPT a good fit can in principle be realized also for $\beta/H_*\gg 100$. However, such large $\beta/H_*$ would require a phase-transition temperature in the sub-MeV range -- in conflict with BBN constraints.} These are properties which naturally occur, for instance, for FOPTs at the end of double-field or chain inflation or for strongly supercooled phase transitions~\cite{Freese:2022qrl,Freese:2023szd}.

The preferred phase-transition temperature falls into the range $T_*\simeq \text{MeV}-\text{GeV}$. It is slightly higher for the FOPTs in the thick-wall regime compared to the thin-wall regime and to thermal phase transitions.
This is because, for thick-wall cases, the fit tries to place the crossover from the steep causality tail in the GW spectrum (with $\Omega_{\text{GW}}\propto f^3$) to the softer below-peak spectrum (with $\Omega_{\text{GW}}\propto f^{0.7}$) within the PTA frequency band, while the peak frequency typically occurs at higher frequency outside the PTA band. In contrast, the GW spectra of the other FOPT scenarios feature a direct transition from the steeply rising causality tail to a falling power-law at the peak frequency (without an intermediate softer power-law). Hence, in these other scenarios the fit prefers to have the peak frequency within the PTA band. For the reasons just described, the preferred peak frequency is higher in the thick-wall scenarios which translates to a larger temperature $T_*$ according to Eq.~\eqref{eq:fpeak}.

\subsection{First-Order Phase Transition vs.\ Supermassive Black Hole Mergers}

So far we considered the SMBH and the FOPT hypotheses separately and identified parameter points best-fitting the observed PTA signal. From a pure goodness-of-fit standpoint both hypotheses (including all subcases) were found to be compatible with the observed GW signal. However, since the SMBH and the FOPT fits contain a different number of dof, the obtained $\chi^2$-values do not directly translate to a preference of one or the other hypothesis. 

In order to gain further insights into the origin of the stochastic GW background, we will perform a frequentist hypothesis test. The advantage of the frequentist approach is that it avoids the assignment of priors to the phase transition parameters. This is in contrast to previously performed Bayesian analyses which have yielded prior-dependent results -- making their interpretation challenging in the light that no strong theoretical guidance on prior choices on the phase transition parameters exists. 

Because SMBH binaries are considered as the default explanation of the PTA signal, we define them as the null-hypothesis in our hypothesis test and a FOPT as the alternative hypothesis. The test statistic is the difference in the best-fit $\chi^2$-values of the FOPT and the SMBH hypothesis,
\begin{equation}
\chi_{\text{FOPT}-\text{SMBH}}^2\equiv\chi^2_\text{PTA}(\text{FOPT})-\chi^2_\text{tot}(\text{SMBH})\,.
\end{equation}
Notice that in a frequentist approach we are not assigning a probability to each hypothesis. Rather, we determine how strongly the null-hypothesis is favored or disfavored against the alternative hypothesis given the experimental outcome. Specifically, we ask the question: assuming the null-hypothesis is true, how likely was an experimental outcome less compatible with the null-hypothesis (i.e.\ how likely was a lower $\chi_{\text{FOPT}-\text{SMBH}}^2$ than the observed one). 

A key input for the hypothesis test is the expected distribution in $\chi_{\text{FOPT}-\text{SMBH}}^2$ under the null-hypothesis, which we need to compare to the observed value. Because the null- and alternative hypothesis rely on distinct GW parameterizations and do not even share the same number of dof, the required $\chi_{\text{FOPT}-\text{SMBH}}^2$-distribution cannot be obtained by analytical methods. Rather, we need to generate a large number of mock data samples under the null-hypothesis and to calculate $\chi_{\text{FOPT}-\text{SMBH}}^2$ for each realization. By repeating this procedure a larger number of times we can obtain the desired distribution. 

In order to produce a mock data set under the null hypothesis (=SMBHs) we generate a random tuple $\{A_{\text{GW}},\gamma_{\text{GW}}\}$ from the model probability distribution extracted from Eq.~\eqref{eq:SMBHparameters}. Because the parameter uncertainties are correlated this is done through the Cholesky decomposition method. In the next step we calculate the expectation value of $\Omega_{\text{GW}}h^2$ for the generated amplitude and power-law index in each PTA bin through Eq.~\eqref{eq:omegahc}. Finally, to obtain the mock measured value in each bin, we allow for a random fluctuation in each bin due to the experimental uncertainties (which are shown in Fig.~\ref{fig:dataextraction}). For each mock data set, we determine the best-fit point for the SMBH hypothesis and for the FOPT hypothesis and the corresponding $\chi$-values -- in complete analogy to what we have done previously for the actual data set. In order to get sufficient statistics we generate 100000 mock data sets under the null hypothesis and extract $\chi_{\text{FOPT}-\text{SMBH}}^2$ from each of them.

\begin{figure}[htp]
\begin{center}
\includegraphics[width=0.85\textwidth]{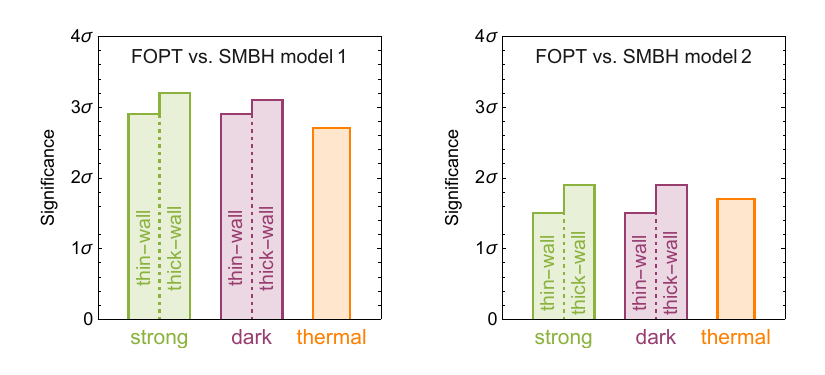}
\end{center}
\vspace{-2mm}
\caption{Statistical significance of different FOPT hypotheses against the SMBH hypothesis. In the left (right) panel SMBH model~1 (model~2) is considered as the null-hypothesis.}
\label{fig:significance}
\end{figure}

We want to consider two versions of the SMBH-hypothesis and five versions of the FOPT hypothesis,
\begin{itemize}
\item null-hypothesis: SMBH model~1/ SMBH model~2,
\item alternative hypothesis: strong FOPT (thin-wall)/ strong FOPT (thick-wall)/ dark FOPT (thin-wall)/ dark FOPT (thick-wall)/ thermal FOPT.
\end{itemize}
Hence, we need to perform ten hypothesis tests in total. We start by testing SMBH model~1 as null-hypothesis against a strong FOPT (thin-wall) as alternative hypothesis. For this case, we extract the observed value of $\chi_{\text{FOPT}-\text{SMBH}}^2=33.9-44.1=-10.2$ from Tab.~\ref{tab:FOPTbest} and Tab.~\ref{tab:SMBHbest}. Within our generated sample of 100000 mock data sets only 388 produce a lower $\chi_{\text{FOPT}-\text{SMBH}}^2$ than the observed one. Assuming the null hypothesis is true such an incompatible outcome thus has a likelihood of 388/100000. Formally, this corresponds to a preference of the alternative hypothesis at 2.9$\sigma$ significance. The same procedure is performed for all versions of the SMBH and FOPT hypotheses. The resulting significances are summarized in Fig.~\ref{fig:significance}.

The performed hypothesis tests reveal a notable preference for the FOPT hypothesis. Compared to the black hole model 1 -- which assumes circular GW-driven binary evolution -- a FOPT is preferred at $\sim 3\sigma$ significance (with slight variation between the different FOPT scenarios). Compared to model~2 -- which considers significant (potentially unrealistic) orbital eccentricities and environmental couplings of the SMBH binaries -- a FOPT is still preferred at $\sim 2\sigma$ significance. However, because the physics of the final parsec in merging SMBHs is not fully understood we refrain from drawing strong conclusions based on small statistical preferences. Nevertheless, it is fair to point out the remarkable consistency of the FOPT hypothesis with the observed PTA signal -- something which would not have been expected if the signal came from SMBH mergers.

Among the different realizations of a FOPT, the largest significance (of slightly above $3\sigma$) is obtained for a strong and for a dark FOPT in the thick-wall regime of vacuum tunneling. These cases allow for the crossover from a steeply rising GW spectrum (=causality tail) to a softly rising spectrum within the PTA frequency band -- a pattern which fits the PTA data particularly well. However, the significance of these thick-wall scenarios is at most $0.5\sigma$ higher than of the other FOPT scenarios. More statistics is, hence, needed to single out any particular FOPT scenario.

\section{Conclusion}\label{sec:conclusion}

In this work we investigated two competing explanations of the GW signal observed by several PTA experiments: a population of SMBH binaries which emit strong GWs at the merger stage of the two black holes and a FOPT in the early Universe. In order to predict the GW spectrum from SMBH binaries we implemented two black hole models proposed in the literature: the conservative model~1 which assumes circular, purely GW-driven binary evolution (during the merging stage relevant for PTA experiments) and the aggressive model~2 which incorporates strong enironmental couplings and excentric orbits and leads to a broader range of possible spectra. Because it is presently unknown, to which extent the effects implemented in model~2 play a role in real SMBH binary systems, the two black hole models reflect the theoretical uncertainties in the GW spectra associated with SMBHs.

For the first-order phase transition, we considered three physically well-motivated subcases: (1) a strong first order phase transition which can, for instance, occur at the end of (thermal) inflation or during the breaking of an extra gauge symmetry in ultraviolet completions of the Standard Model of particle physics (if there is supercooling), (2) a dark phase transition possibly associated with the production of dark matter, (3) a thermal phase transition reflecting the standard case of a phase transition occurring during radiation domination. By implementing these three subcases -- which capture a wide range of possible spectra, ranging from bubble-collision-dominated GW production (strong and dark phase transition) to sound-wave-dominated GW production (thermal phase transition) -- we aimed to also gain insights into the nature of the FOPT preferred by the PTA data (if any).

In the first step we investigated the general compatibility of the SMBH and FOPT hypotheses with the observed PTA data by means of a goodness-of-fit test. We found that both hypotheses (including all subcases) yield best-fit values of $\chi^2/\text{dof}\lesssim 1$ indicating agreement with the data at the $1\sigma$-level.
The large observed GW amplitude only matches the very upper end of the prediction from SMBHs (see Fig.~\ref{fig:bhfit}, where we included the full collection of NANOGrav, PPTA and EPTA data sets). Furthermore, within the SMBH models a somewhat better fit is obtained for model~2 suggesting that, if the PTA signal is of the standard astrophysical origin, environmental effects and/or eccentricities could indeed play a role in SMBH binary systems (at the separation scales resolved in PTA experiments) -- consistent with previous findings in the literature~\cite{NANOGrav:2023hvm,EPTA:2023xxk,Ellis:2023oxs}.

In order to assess a possible preference for a FOPT vs.\ SMBHs we performed a frequentist hypothesis test, where we defined SMBH mergers as the null-hypothesis and a FOPT as the alternative hypothesis. Compared to previous studies in the literature our approach profited from including the full collection of NANOGrav, PPTA and EPTA data sets (instead of focusing on a single experiment). Furthermore, due to our frequentist approach we were able to avoid any dependence of our results on priors on the phase transition parameters (which complicated the interpretation of the previous analyses which all followed a Bayesian approach).

Since the two hypotheses do not share the same degrees of freedom, the hypothesis test required the generation of a large sample of mock experimental outcomes under the null-hypothesis.  By comparing the expected distribution of $\chi^2$-difference between the null- and the alternative hypothesis (as extracted from the mock sample) with the observed $\chi^2$-difference we were able to determine the statistical preference for the alternative hypothesis. We found a notable preference for the FOPT hypothesis at $\sim 3\sigma$ significance compared to black hole model~1 and at $\sim 2\sigma$ significance compared to model~2.  Due to the lack of understanding of the physics of the final parsec in merging SMBH systems, and due to possible systematics in the GW spectra from FOPTs it is too early to draw any strong conclusions from these results. Nevertheless, one can say that the compatibility of a FOPT with the PTA signal is remarkably good. More robust evidence for the FOPT hypothesis could arise, in the future, if a clear peak in the GW background can be established (because no such peak is expected within the SMBH hypothesis). 

 Among the different realizations of FOPTs, the largest significance (of slightly above $3\sigma$) is obtained for a strong and for a dark FOPT. These cases allow for the crossover from a steeply rising GW spectrum (=causality tail) to a softly rising spectrum within the PTA frequency band -- a pattern which fits the PTA data particularly well. The peak frequency would fall slightly above the PTA-measured range in this case. However, the significances of all FOPT scenarios fall within $0.5\sigma$ of one another,  suggesting the need of more statistics and ideally an extension of the PTA frequency band (see e.g.~\cite{DeRocco:2022irl,DeRocco:2023qae}).

Future more precise measurements of the GW background, for instance with the Square Kilometre Array~\cite{Carilli:2004nx,Janssen:2014dka,Weltman:2018zrl}, as well as the possible resolution of individual GW sources will provide further insights into the origin of the signal.

\section{Acknowledgments}
We thank Ben Lehmann for very helpful conversations at the beginning of this project. Furthermore, we would like to thank Kimberly Boddy and Barmak Shams Es Haghi for very useful comments. K.F. and M.W. are grateful for support from the Jeff and Gail Kodosky Endowed Chair in Physics at the Univ. of Texas, Austin. K.F. and M.W. acknowledge funding from the U.S. Department of Energy, Office of Science, Office of High Energy Physics program under Award Number DE-SC0022021. K.F. acknowledges support by the Vetenskapsradet (Swedish Research Council) through contract No. 638- 2013-8993 and the Oskar Klein Centre for Cosmoparticle Physics at Stockholm University.
\bibliography{ptas}
\bibliographystyle{h-physrev}

\end{document}